\documentclass{article}

\usepackage{amsmath}
\usepackage{arxiv}

\usepackage[utf8]{inputenc} 
\usepackage[T1]{fontenc}    
\usepackage{hyperref}       
\usepackage{url}            
\usepackage{booktabs}       
\usepackage{amsfonts}       
\usepackage{nicefrac}       
\usepackage{microtype}      
\usepackage{lipsum}
\usepackage{graphicx}
\graphicspath{ {./images/} }
\usepackage[numbers,sort&compress]{natbib}
\usepackage{color,soul}
\usepackage{natbib}
\usepackage[flushleft]{threeparttable}
\usepackage{booktabs}
\usepackage{caption}
\usepackage{subcaption}
\usepackage{tabularx}
\usepackage{multirow}
\usepackage{authblk} 

\hypersetup{
  colorlinks,
  citecolor=blue,
  linkcolor=blue,
  urlcolor=black}

\title{LoRA-BERT: a Natural Language Processing Model for Robust and Accurate Prediction of long non-coding RNAs}

\author[1]{Nicholas Jeon}
\author[1,2]{Xiaoning Qian}
\author[3]{Lamin SaidyKhan}
\author[3,4]{Paul de Figueiredo}
\author[1,2,*]{Byung-Jun Yoon}
\affil[1]{Department of Electrical and Computer Engineering, Texas A\&M University}
\affil[2]{Computing and Data Sciences, Brookhaven National Laboratory, }
\affil[3]{Christopher S. Bond Life Sciences Center, Department of Molecular Microbiology and Immunology, The University of Missouri}
\affil[4]{Department of Veterinary Pathobiology, Department of Chemical and Biomedical Engineering, University of Missouri School of Medicine}
\affil[*]{Corresponding author. \href{email:bjyoon@tamu.edu}{bjyoon@tamu.edu}}

\begin{document}
\maketitle

\begin{abstract}
Long non-coding RNAs (lncRNAs) serve as crucial regulators in numerous biological processes. Although they share sequence similarities with messenger RNAs (mRNAs), lncRNAs perform entirely different roles, providing new avenues for biological research. The emergence of next-generation sequencing technologies has greatly advanced the detection and identification of lncRNA transcripts and deep learning-based approaches have been introduced to classify long non-coding RNAs (lncRNAs). These advanced methods have significantly enhanced the efficiency of identifying lncRNAs. However, many of these methods are devoid of robustness and accuracy due to the extended length of the sequences involved. To tackle this issue, we have introduced a novel pre-trained bidirectional encoder representation called LoRA-BERT. LoRA-BERT is designed to capture the importance of nucleotide-level information during sequence classification, leading to more robust and satisfactory outcomes. In a comprehensive comparison with commonly used sequence prediction tools, we have demonstrated that LoRA-BERT outperforms them in terms of accuracy and efficiency. Our results indicate that, when utilizing the transformer model, LoRA-BERT achieves state-of-the-art performance in predicting both lncRNAs and mRNAs for human and mouse species. Through the utilization of LoRA-BERT, we acquire valuable insights into the traits of lncRNAs and mRNAs, offering the potential to aid in the comprehension and detection of diseases linked to lncRNAs in humans.
\end{abstract}


\section{Introduction}
Long non-coding ribonucleic acids~(lncRNAs), a class of non-coding RNAs with $>$200 nucleotides that have no or limited protein-coding capacity, have attracted significant attention among the molecular biology community, due to the discovery of their involvement in important cellular and physiological functions that were previously unappreciated \cite{guttman2013ribosome,derrien2012gencode,harrow2012gencode,ma2023contribution}. Due to the high discrepancy between the number of protein-coding genes and total transcripts identified in mammalian transcriptomes, the chances of discovering novel functions for non-translated RNAs, including lncRNAs, is high \cite{djebali2012landscape,pennisi2012encode,yang2014oncogenic}. In fact, there has been an exponential increase in the number of reports highlighting the dynamic expression and biological functions of lncRNAs \cite{ma2023contribution, fernandes2019long}. With the advent of high—throughput and deep sequencing tools, evidence of lncRNA participation in diverse biological processes, including in non-transcriptional regulation \cite{rinn2012genome,bhartiya2012conceptual,lu2013computational} and genetic regulation (such as chromosome structure modulation, transcription, splicing, messenger RNA (mRNA) stability, mRNA availability, and post-translational modification) has been revealed \cite{fernandes2019long}. In addition, numerous lines of evidence indicate that lncRNAs are closely linked to a broad spectrum of human health conditions, such as lung cancer \cite{shi2015critical}, central nervous system disorders \cite{ng2013long}, and heart diseases \cite{congrains2012genetic}. Despite the massive sequence data that includes analyses of information about the expression of lncRNAs, the accuracy of defining transcriptional units using current sequencing tools is not reliable, making the identification and functional characterization of lncRNAs challenging. Therefore, there is a great need for the development of new accurate, and robust algorithms (such as natural language processing, NLP) for understanding the structural and mechanistic features of lncRNAs. 

Various databases and experiments have been established to support studies on lncRNAs from their identification to annotation \cite{xu2017comprehensive}. Numerous methods and tools have emerged, employing techniques like Support Vector Machine (SVM) \cite{li2014plek,kang2017cpc2,han2019lncfinder} and machine learning \cite{fan2020lncrna_mdeep} to predict lncRNA. More recent approaches have used Convolutional Neural Network (CNN)~\cite{zhang2020plant,zou2019primer} and Long Short Term Memory (LSTM)~\cite{hochreiter1997long} models to classify the RNA sequence category. Additionally, Hybrid methods that combine the strengths of CNN and LSTM model architectures have also been proposed \cite{liu2019prediction}.

However, it's important to note that both CNN and RNN architectures have limitations when it comes to handling long-range contexts, as seen in lncRNAs and mRNAs. CNNs often struggle to capture semantic dependencies in sequences of substantial length due to their limited capacity to extract local features based on filter size. On the other hand, RNN models, while capable of learning long-term dependencies, frequently encounter challenges related to vanishing gradients and processing efficiency when handling sequences in a sequential manner.

To overcome the limitations mentioned above, we have taken the idea from the Bidirectional Encoder Representations from Transformers (BERT) model by \cite{devlin2018bert} and adapted it for the context of genomics. Utilizing the BERT model, we have developed LoRA-BERT (\textbf{Lo}ng noncoding \textbf{R}N\textbf{A}-\textbf{BERT}), which employs the Transformer architecture that relies on attention mechanisms and has consistently achieved great performance in most NLP tasks, as demonstrated by \cite{ji2021dnabert}. LoRA-BERT effectively addresses the challenges outlined above by effectively capturing context information from the entire input sequence on a global scale using an attention mechanism. LoRA-BERT reveals critical subregions within sequences and identifies potential relationships between different RNA sequences. Our approach primarily leverages features like $k$-mer and Open Reading Frame (ORF) to assist the model in classifying the input sequence. With the help of this NLP-inspired model, LoRA-BERT can proficiently distinguish between lncRNAs and mRNAs originating from both human and mouse species. 

\section{Materials and Methods}

\subsection{Data Description}

\begin{table*}
    \centering
        \begin{threeparttable}
          \caption{Description of four different transcripts of human and mouse for lncRNA and mRNA prediction}
	\label{table:dataset}
	
	\begin{tabularx}{\linewidth}{XXXXXXX}
		\toprule
		\textbf{Dataset} & \textbf{Database} & \textbf{Transcript} & \textbf{Size} & \textbf{Min length} & \textbf{Max length} & \textbf{Mean Length}\\
		\midrule
		\multirow{2}{*}{Human}     & GRCh38.cds & mRNA &59,419 &100 &107,976 &1,363\\
            &GRCh38.ncrna & lncRNA &59,419 &100 &347,561 &1,293\\ 
		\midrule
		\multirow{2}{*}{Mouse} & GRCm39.cds  & mRNA &22,583 &100 &106,173 &1,431\\
            &GRCm39.ncrna & lncRNA &22,583 &101 &93,147 &1,210\\
		\bottomrule
	\end{tabularx}
         \begin{tablenotes}
      \small
      \item The origin of our data is attributed to the Ensembl databases. The size parameter indicates the number of sequences present within the transcript, while min, max, and mean length represent the smallest, largest, and average lengths of sequences within each transcript, measured by the length of the sequence. It's important to note that we excluded any sequences with a length of less than 100 nucleotides from the transcripts. 
    \end{tablenotes}
  \end{threeparttable}
\end{table*}

As research has advanced, the identification of lncRNAs and mRNAs has emerged, resulting in the development of comprehensive transcriptional RNA databases. These databases have supplied specific datasets that facilitate the recognition of lncRNAs and mRNAs. Ensembl, functioning as a genome browser tailored to vertebrate genomes, lends support to research efforts across the domains of comparative genomics, evolutionary studies, sequence variation analysis, and the investigation of transcriptional regulatory mechanisms. Ensembl is equipped to annotate genes, perform complex multiple alignments, forecast regulatory functions, and amass data related to diseases. It encompasses a vast, unified, non-repetitive, and meticulously annotated dataset that encompasses a wide range of information, including mRNA and lncRNA sequences, among other data types. Additionally, Ensembl provides data pertaining to both human and mouse species, enhancing its value as a comprehensive resource for biological research.

To train and test the performance of the LoRA-BERT model, we collected two types of species, the human and mouse from the Ensembl database. After collecting the datasets, we removed the sequences less than 100 nts in length and also eliminated any unannotated sequences from the transcript. In the human case, 59,419 sequences were collected for each lncRNA and mRNA. Likewise, 22,583 sequences were collected for the mouse case. Detailed data information is presented in the Table~\ref{table:dataset}.

\subsection{Model Architecture}

\begin{figure}
        \captionsetup{width=0.7\textwidth}
	\centering
	\includegraphics[scale=0.8]{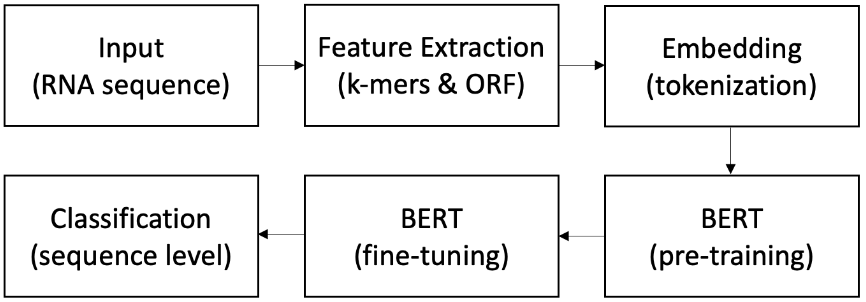}
	\caption{The model's schematic diagram. First, we use feature extraction to partition each input RNA sequence. Subsequently, with the complete set of $k$-mer sequences, the tokenization layer is responsible for acquiring embedding vectors for all these $k$-mers. This layer converts all the $k$-mer sequences into a matrix in the continuous vector space. The model consists of 12 BERT layers with multi-head self-attention, two layer norm, and feed-forward. The last layer, classification layer, calculates the probability and the result of the classification for the input sequence, determining whether it falls into a positive or negative class. }
	\label{fig:fig1}
\end{figure}

In our effort to classify between lncRNAs and mRNAs, we employed two classification features: the $k$-mer pattern and ORF. For the task of identifying full sequences, our classification relied solely on the $k$-mer pattern as the exclusive feature. However, when it came to detecting partial sequences, we incorporated the ORF feature before applying the $k$-mer pattern. The outcome of the $k$-mer pattern analysis, which corresponds to an RNA sequence, is denoted by either "0" for mRNAs or "1" for lncRNAs. The main objective of this paper is to develop a classifier capable of predicting whether a given sequence is lncRNAs or mRNAs. The entire model framework is illustrated in Figure~\ref{fig:fig1}.

\subsection{K-mer Pattern}

We focused exclusively on utilizing the $k$-mer representation as the classification feature between lncRNAs and mRNAs. To accomplish this, we implemented a procedure that involved intercepting RNA sequences, which consisted of A, C, T, and G nucleotides, concatenating the next nucleotide with a step size of $s$ and a length size of $k$. Each segment, comprising $k$ nucleotides, was denoted as a $k$-mer pattern. Overall, the combination of all the $k$-mer segments extracted from an input sequence formed a $k$-mer sequence with a length of L. For a specific value of $k$, it is possible to generate $4^k$ distinct $k$-mer patterns. For instance, if $k=3$, there would be 64 possible patterns such as "AAA," "AAC," "AAT," and so forth.

Subsequently, we mapped these $k$-mer patterns to numerical values based on a predefined dictionary. These numerical values allowed us to create initial input features, where each $k$-mer pattern corresponded to a unique index. For instance, "AAA" corresponds to index 1, while "AAC" corresponds to index 2. Ultimately, each of the $k$-mer segments was linked to a set represented as $\mathbb{Z}=[1,2,3,...,4^k]$.

\subsection{Open Reading Frame}
The ORF is a genetic segment with potential for translation and holds a central role in the context of identifying lncRNAs. This prominence is attributed to the feature distinction between ORF lengths in mRNAs and lncRNAs, where mRNAs typically exhibit more extended ORFs. Consequently, the identification of lncRNAs relies significantly on the exploration and utilization of diverse ORF-related features. Unlike protein-coding genes, lncRNAs present a distinct pattern of random distribution for start and stop codons involved in the translation process. Consequently, it is exceedingly rare for the ORF length in lncRNAs to exceed 100 nucleotides. This rarity underscores the fundamental role of ORF length as a central characteristic for distinguishing between lncRNAs and mRNAs.

In the quest for distinguishing the portion sequences, the $k$-mer feature was not able to complete the task, as the arrangement of $k$-mer patterns within the $k$-mer sequence was completely altered. To address this challenge, we also gathered the longest ORF within the lncRNAs and mRNAs from the corresponding sequences in the transcript. Consequently, by comparing the ORFs of the partial sequences with those we extracted from the transcript for lncRNAs and mRNAs, it became feasible to determine whether the partial sequence belonged to lncRNAs or mRNAs.

\subsection{Tokenization}
Rather than considering each sequence as one singular token, we tokenized RNA sequences using the $k$-mer representation, a widely accepted method for analyzing RNA sequences. This $k$-mer sequence improves the information associated with the context for each deoxynucleotide base by combining nucleotide counts through concatenation. In our experiments, we explicitly configured the value of $k$ to be 3. The vocabulary includes all possible permutations of the $k$-mers, along with five distinct tokens: [CLS] representing the classification token, [SEP] denoting the separation token, [UKN] representing the unknown token, [PAD] indicating the padding token, and [MASK] signifying the masked token. Consequently, our model's vocabulary comprises a total of $4^k+5$ tokens.

\subsection{LoRA-BERT Model}

\begin{figure*}
     \centering
     \captionsetup[subfigure]{font={bf,small}, skip=-1pt, margin=0cm, singlelinecheck=false}
     \begin{subfigure}{0.6\textwidth}
         \centering
         \caption{}
         \includegraphics[width=\textwidth]{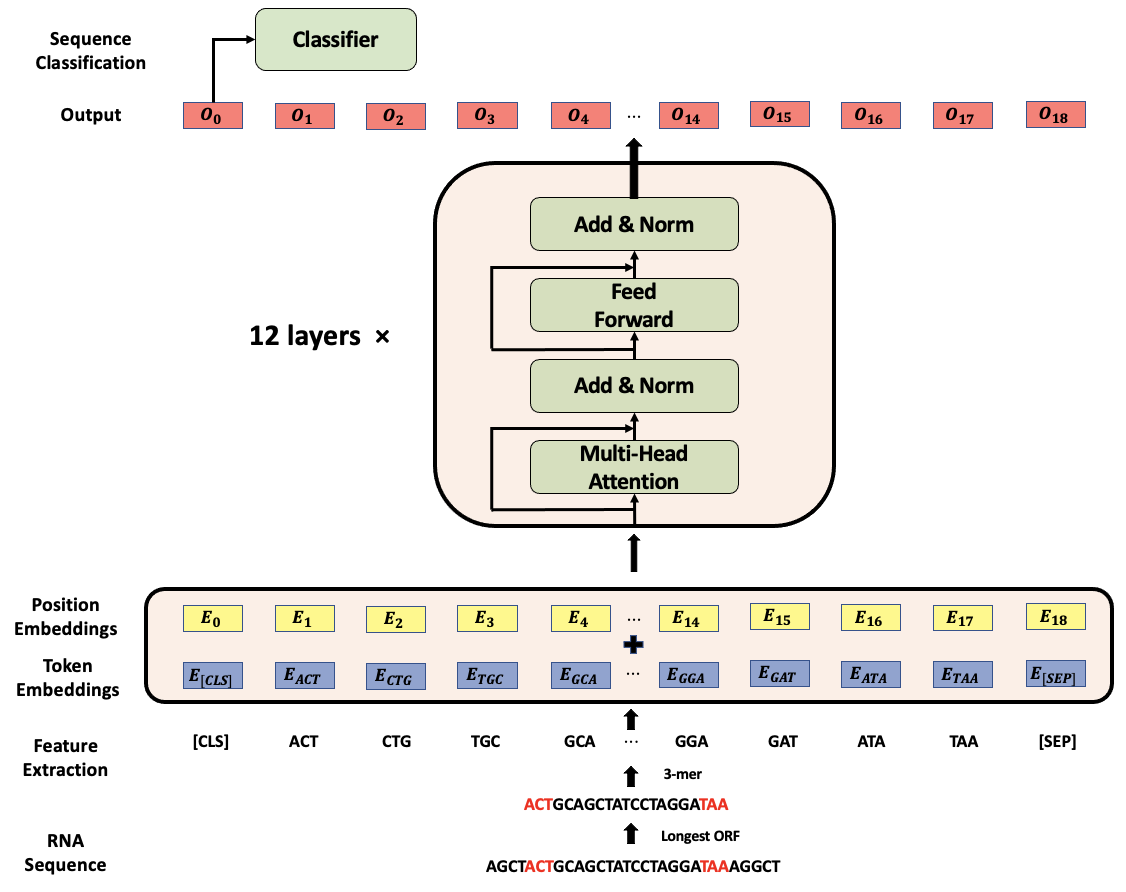}
         \label{fig:a}
     \end{subfigure}
     \hfill
     \begin{subfigure}{0.276\textwidth}
         \centering
         \caption{}
         \includegraphics[width=\textwidth]{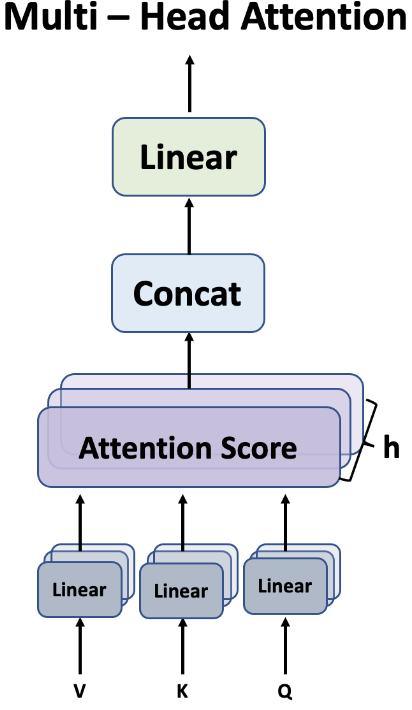}
         \label{fig:b}
     \end{subfigure}
        \caption{LoRA-BERT model architecture: (a) LoRA-BERT uses a tokenized feature extracted sequence as input, which contains classification and separate tokens. The tokenized sequence goes through embedding layers and passes 12 transformer layers. We utilized the initial output from the last hidden states for sequence-level classification. (b) The Multi-Head Attention architecture comprises multiple attention score layers operating simultaneously in parallel.}
        \label{fig:myfigure}
\end{figure*}
    
BERT, a transformer-driven model for contextual language representation, has achieved remarkable success in numerous natural language processing (NLP) tasks, surpassing human-level performance. It introduces a distinctive approach involving pre-training on extensive datasets to establish a broad understanding of language, followed by fine-tuning for specific tasks with minimal architectural adjustments. In our project, we adhere to the same training process as DNABERT model \citep{ji2021dnabert}, but we have chosen to employ the BertForSequenceClassification model instead of the standard BERT model, as our primary objective is to classify input sequences.

Our model begins by receiving a collection of sequences, which are encoded as $k$-mer tokens (Fig~\ref{fig:a}). To represent each sequence, we transform these tokens into numerical vectors, forming a matrix denoted as $M$. In a formal sense, our model captures the multi-head attention mechanism on matrix $M$ (Fig~\ref{fig:b}) to capture contextual information:

\begin{equation}
	MultiHead(Q,K,V) = Concat(head_1,...,head_h)W^O,
\end{equation}
where
\begin{equation}
    \begin{aligned}
	head_i &= Attention Score(QW_{i}^{Q},KW_{i}^{K},VW_{i}^{V}) \\
               &= softmax(\frac{QW_{i}^{Q}(KW_{i}^{K})^T}{\sqrt{d_k}})VW_{i}^{V}.
    \end{aligned}
\end{equation}
The attention score takes as input a set of queries $Q \in \mathbb{R}^{L \times d_k}$, keys $K \in \mathbb{R}^{L \times d_k}$, and values $V \in \mathbb{R}^{L \times d_v}$, where $L$ is the sequence length, $d_k$ and $d_v$ are the hidden dimensionality for queries/keys and values respectively, $W_{i}^{Q}, W_{i}^{K}, W_{i}^{V},$ and $ W_{i}^{O}$ are the learnable parameters. In our model, we assign queries, keys, and values as the matrix M and calculate the attention score \cite{vaswani2017attention}.

Much like BERT, our model also embraces a pre-training and fine-tuning approach. However, we made some alterations to the initial pre-training procedure compared to the original BERT setup. Notably, we eliminated the mask and next sentence prediction components and focused on the sequence classification. During the pre-training phase, our model acquires fundamental knowledge of sentence structure and sequence patterns, utilizing sequences of varying lengths ranging from 100 to 1,536, which are extracted from human and mouse transcripts. In our pre-training process, we employed a cross-entropy loss function denoted as $L = \sum_{i=0}^{N} -y_{i}^{'}log(y_i)$, where $y_{i}^{'}$ and $y_{i}$ represent the actual label and predicted probabilities for each of the N different categories.

\section{Results}

\subsection{Pre-training and Fine-tuning}
Continuing from the earlier research \cite{devlin2018bert, liu2019roberta, yang2019xlnet}, LoRA-BERT processes input sequences with a maximum length of 1,536. As shown in Figure~\ref{fig:a}, when working with an RNA sequence, we break it down into a sequence of $k$-mers. We append a special token "classification" at the sequence's starting point, signifying the entire sequence, and append another special token "separate" at the end, indicating the sequence's end. During the pre-training phase, we removed extraneous tasks such as masking and next-sentence prediction, instead concentrating on the task of sequence classification. We conducted pre-training for 6,340 steps with a batch size of 64. Initially, the loss decreased from 0.211 to 0.149 over the first 6,000 steps. We paused the training process and saved the best performance because, at the 6,000 steps, the model had overfitting. We employed the same model architecture as the DNABERT base, which consists of 12 transformer layers, each featuring 768 hidden units and 12 attention heads. Our model possessed 86,906,114 trainable parameters and was trained using an NVIDIA A100 40GB GPU.

Utilizing a pre-trained model offers notable advantages. It lowers computational expenses and grants access to cutting-edge models without the need to train one from the ground up. Transformers offer access to an extensive array of pre-trained models across various tasks. In each subsequent application, we initiated the process using the pre-trained model as our starting point. we have built and fine-tuned with different datasets that have not been used in the pre-training part. We applied the same training techniques in all the scenarios, wherein the learning rate initially underwent a linear warm-up to reach its maximum value, and subsequently, it was linearly reduced to approach nearly 0. In this part, the loss decreased from 0.182 to 0.128. If the users wish to train with their custom datasets, they can employ the pre-trained model we have developed.

\subsection{Performance for full-sequence identification }

\begin{table}
    \centering
          \caption{Comprehensive outcomes of LoRA-BERT's performance on each dataset, including loss and accuracy on both training and test datasets.}
	\label{table:eval}
	\begin{tabularx}{\linewidth}{XXXXX}
		\toprule
		\multirow{2}{*}{\textbf{Dataset}} & \textbf{Train} & \textbf{Test} & \textbf{Train} & \textbf{Test} \\
                                              & \textbf{Loss} & \textbf{Loss} & \textbf{Accuracy} & \textbf{Accuracy} \\
		\midrule
		Human     & 0.169 & 0.143 &0.955 &0.985 \\ 
		\midrule
		Mouse & 0.130  & 0.164 &0.965 &0.977 \\
		\bottomrule
	\end{tabularx}
\end{table}

\begin{table*}
    \centering
        \begin{threeparttable}
          \caption{The comparison with four different models in lncRNA and mRNA prediction for full-sequence transcript.}
	\label{table:full_comaparison}
	
	\begin{tabularx}{\linewidth}{XXXXXXX}
		\toprule
		\textbf{Dataset} & \textbf{Methods} & \textbf{Precision} & \textbf{Recall} & \textbf{Accuracy} & \textbf{F1-score} & \textbf{auROC} \\
		\midrule
		& LncFinder & 0.831& 0.935& 0.872& 0.879& 0.927\\
            &Mdeep & 0.969& 0.948& 0.959& 0.958& 0.989\\ 
      Human &PLEK & 0.707& 0.919& 0.764& 0.799& 0.821\\ 
            &CPC2 & 0.789& 0.924& 0.838& 0.851& 0.911\\ 
            &\textbf{\mbox{LoRA-BERT}} & \textbf{0.980}& \textbf{0.989}& \textbf{0.985}& \textbf{0.985}& \textbf{0.998}\\ 
		\midrule
		& LncFinder  & 0.972& 0.953& 0.963& 0.962& 0.984\\
            &Mdeep & \textbf{0.974}& 0.907& 0.942& 0.940& 0.987\\ 
      Mouse &PLEK & 0.694& 0.888& 0.745& 0.779& 0.797\\ 
            &CPC2 & 0.830& 0.929& 0.869& 0.877& 0.934\\ 
            &\textbf{\mbox{LoRA-BERT}} & 0.968& \textbf{0.988}& \textbf{0.977}& \textbf{0.978}& \textbf{0.997}\\ 
		\bottomrule
	\end{tabularx}
  \end{threeparttable}
\end{table*}

\begin{figure*}
    \centering
    \captionsetup[subfigure]{font={bf,small}, skip=-1pt, margin=0cm, singlelinecheck=false}
    \begin{subfigure}{0.45\textwidth}
        \centering
        \caption{} 
        \includegraphics[width=\textwidth]{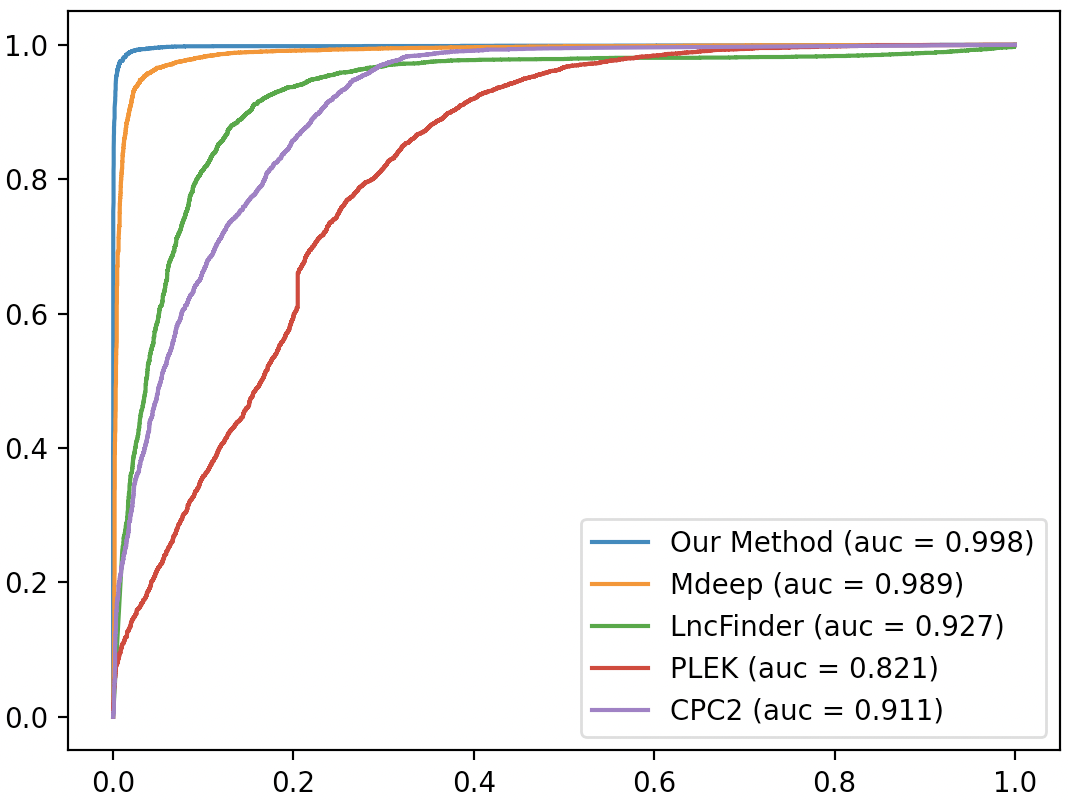}
        \label{fig:hu_full}
    \end{subfigure}
    \hfill
    \begin{subfigure}{0.45\textwidth}
        \centering
        \caption{} 
        \includegraphics[width=\textwidth]{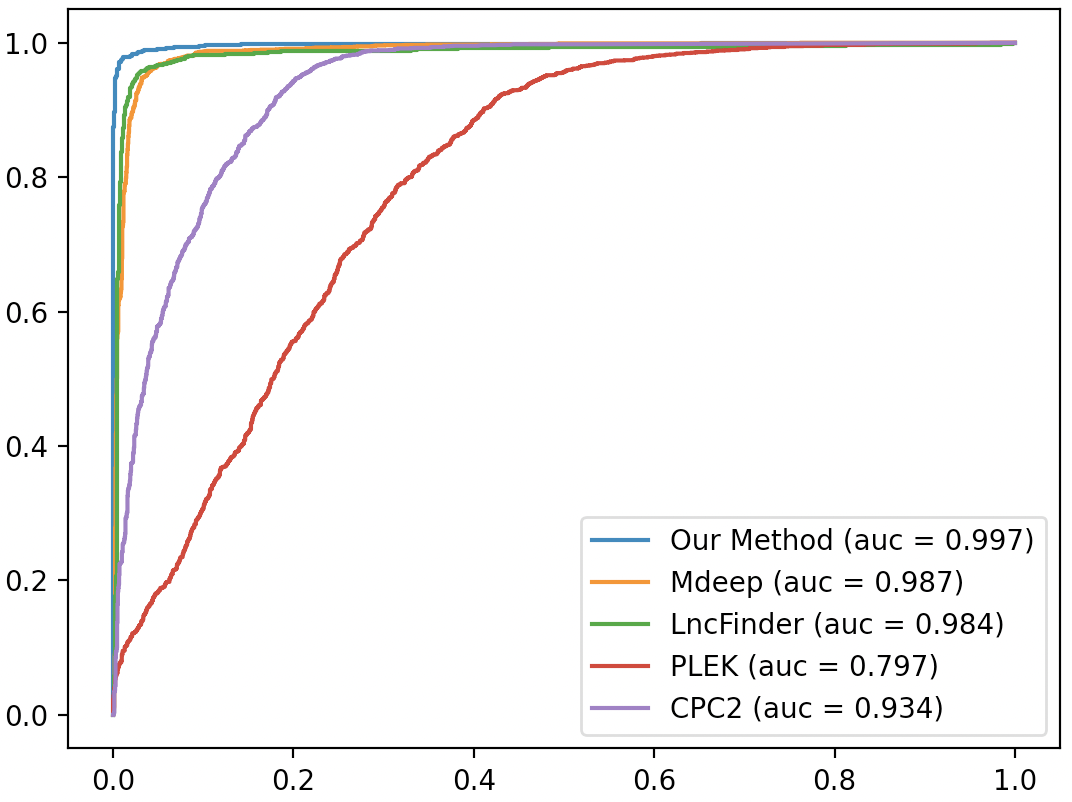}
        \label{fig:mo_full}
    \end{subfigure}
        \caption{The comparison with different models for ROC curve of (a) human and (b) mouse, with TPR as the vertical axis and FPR as the horizontal axis.}
        \label{fig:full_roc}
\end{figure*}

Initially, we presented the evaluation outcomes of our model on each dataset in Table~\ref{table:eval} and listed the loss and accuracy in training and testing sets for the full-sequence transcripts. The results for both the training set and the test set exhibit a resemblance, highlighting the efficacy of the early termination strategy in mitigating overfitting. Significantly, the most precise prediction was obtained for the human dataset, achieving an impressive accuracy score of 98.5\%.

Next, We conducted a comparison between our model and several baseline methods to evaluate their performance, including LncFinder \citep{han2019lncfinder}, Mdeep \citep{fan2020lncrna_mdeep}, PLEK \citep{li2014plek}, CPC2 \citep{kang2017cpc2}. For LncFinder, we used the R program to implement and chose the hyperparameter set advised by Han, which produced the best-recorded results. For CPC2, we directly accessed the website mentioned in the article and evaluated it using our test dataset. For PLEK, we used Python code and followed the same procedure mentioned by Aimin. For Mdeep, we followed the instructions in the GitHub provided in the article. To evaluate the performance, we used metrics including precision, recall, accuracy, F1-score, and auROC. The precise calculation formulas are outlined as follows::
\begin{equation}
	\mbox{Precision} = \frac{\mbox{TP}}{\mbox{TP} + \mbox{FP}}
\end{equation}

\begin{equation}
	\mbox{Recall} = \frac{\mbox{TP}}{\mbox{TP} + \mbox{FN}}
\end{equation}

\begin{equation}
	\mbox{Accuracy} = \frac{\mbox{TP} + \mbox{TN}}{\mbox{TP} + \mbox{TN} + \mbox{FP} + \mbox{FN}}
\end{equation}

\begin{equation}
	\mbox{F1-score} = 2\times\frac{\mbox{Precision} \times \mbox{Recall}}{\mbox{Precision} + \mbox{Recall}}
\end{equation}

\begin{equation}
	\mbox{TruePositiveRate} = \frac{\mbox{TP}}{\mbox{TP} + \mbox{FN}}
\end{equation}

\begin{equation}
	\mbox{FalsePositiveRate} = \frac{\mbox{FP}}{\mbox{FP} + \mbox{TN}}
\end{equation}

where TP, TN, FP, and FN represent true positive, true negative, false positive, and false negative, respectively. AuROC, which stands for the area under the receiver operating characteristic~(ROC) curve, was determined by creating the ROC curve with the "true positive rate" on the vertical axis and the "false positive rate" on the horizontal axis and then calculating the area beneath this curve.

As results from the comparison presented in Table~\ref{table:full_comaparison} and Figure~\ref{fig:full_roc}, our approach outperformed the other baseline methods. Especially on precision, recall, accuracy, f1-score, and auROC, our model received better results than any other existing model with 98.0\%, 98.9\%, 98.5\%, 98.5\%, and 99.8\% respectively in human species. On the other hand, while Mdeep exhibited higher precision than our model, the rest of our results achieved the highest performance for the recall, accuracy, f1-score, and auROC in mouse species. These results highlight that the NLP model with feature extraction outperformed the SVM model that relied on manually extracting $k$-mer features. This demonstrates the robustness of our model, which can be effectively pre-trained and fine-tuned for accurate predictions, not only for short sequence lengths but also for longer sequences like lncRNA and mRNA in the human and mouse genome, relying solely on nearby $k$-mer patterns.

\subsection{Performance for partial-sequence identification }

Lastly, we evaluate the performance of partial sequence transcript. The origin of the datasets is the same as the full sequence transcript but we have removed the front and last part with the portion of 10\%, 20\%, 30\%, and 40\% respectively. For example, if the remaining portion is 90\% then we removed 5\% of the front region and 5\% of the last region and kept the middle part as the partial sequence and tested on LoRA-BERT.

\begin{figure*}
    \centering
    \captionsetup[subfigure]{font={bf,small}, skip=-1pt, margin=0cm, singlelinecheck=false}
    \begin{subfigure}{0.4\textwidth}
        \centering
        \caption{} 
        \includegraphics[width=\textwidth]{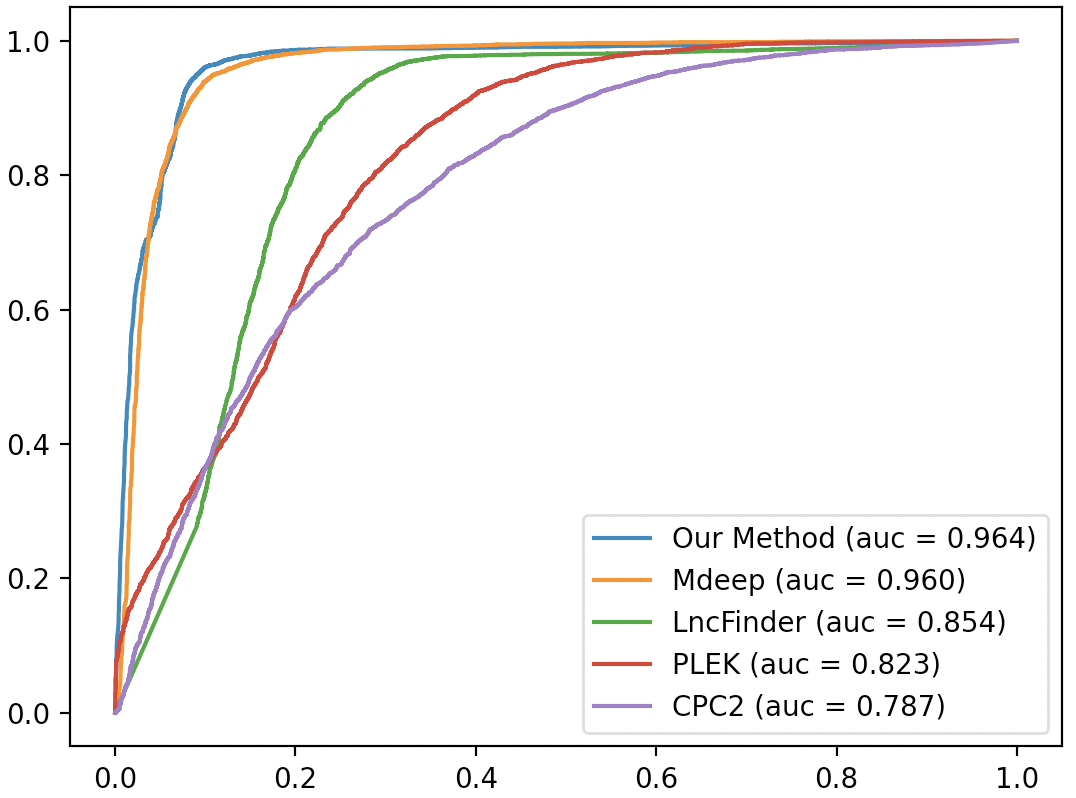}
        \label{fig:hu_90}
    \end{subfigure}
    \begin{subfigure}{0.4\textwidth}
        \centering
        \caption{} 
        \includegraphics[width=\textwidth]{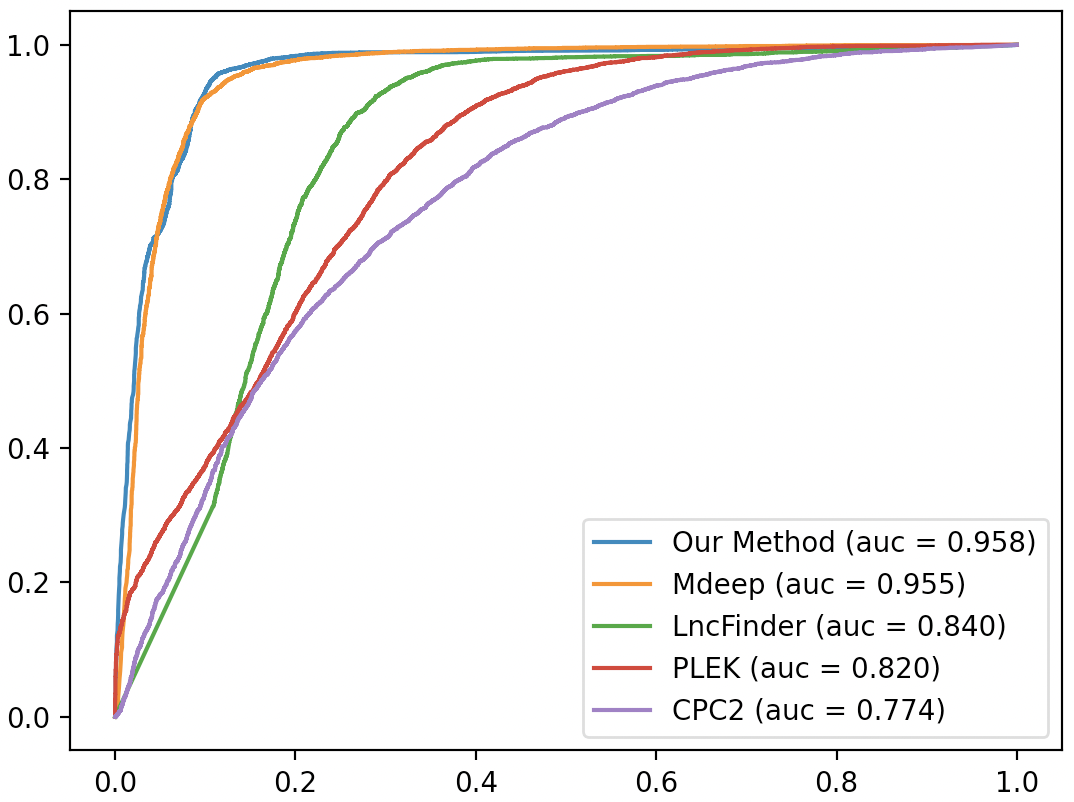}
        \label{fig:hu_80}
    \end{subfigure}
    \\
    \begin{subfigure}{0.4\textwidth}
        \centering
        \caption{} 
        \includegraphics[width=\textwidth]{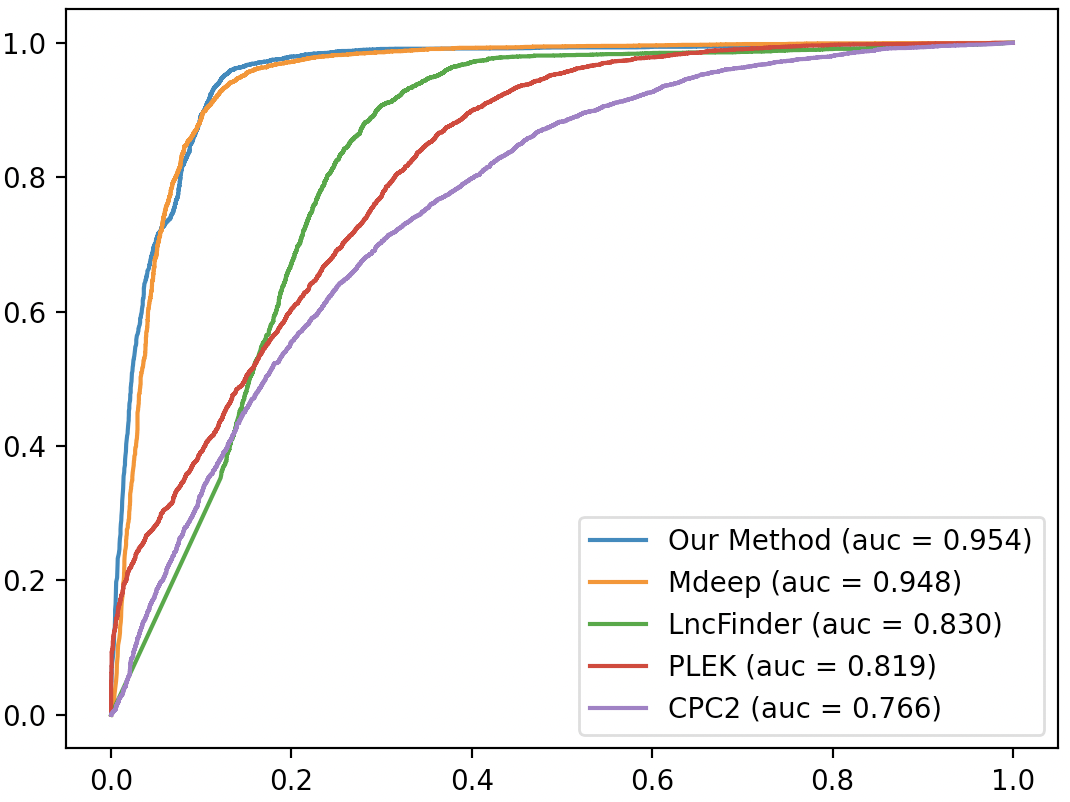}
        \label{fig:hu_70}
    \end{subfigure}
    \begin{subfigure}{0.4\textwidth}
        \centering
        \caption{} 
        \includegraphics[width=\textwidth]{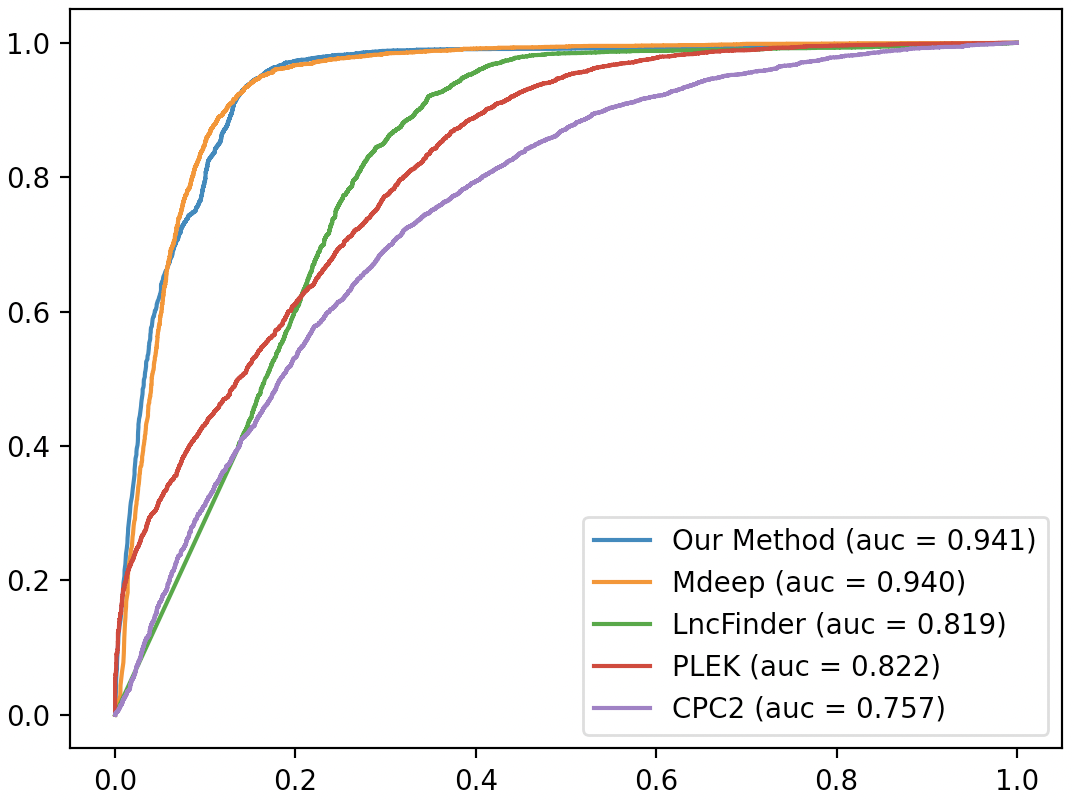}
        \label{fig:hu_60}
    \end{subfigure}
        \caption{The comparison with different models for ROC curve for different partial ratios for human species. (a) represent 90\% of partial sequence ROC curve, (b) represent 80\% of partial sequence ROC, (c) represent 70\% of partial sequence ROC curve, and (d) represent 60\% of partial sequence ROC curve.}
        \label{fig:hu_partial_roc}
\end{figure*}

\begin{figure*}
    \centering
    \captionsetup[subfigure]{font={bf,small}, skip=-1pt, margin=0cm, singlelinecheck=false}
    \begin{subfigure}{0.4\textwidth}
        \centering
        \caption{} 
        \includegraphics[width=\textwidth]{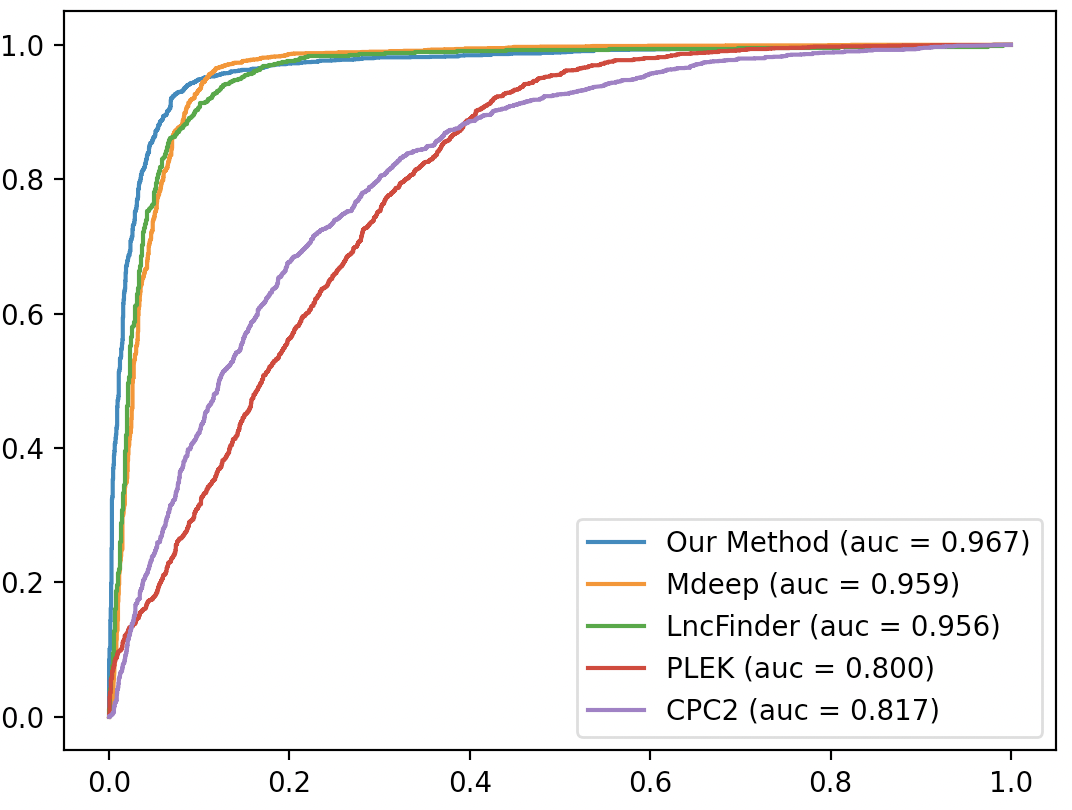}
        \label{fig:mo_90}
    \end{subfigure}
    \begin{subfigure}{0.4\textwidth}
        \centering
        \caption{} 
        \includegraphics[width=\textwidth]{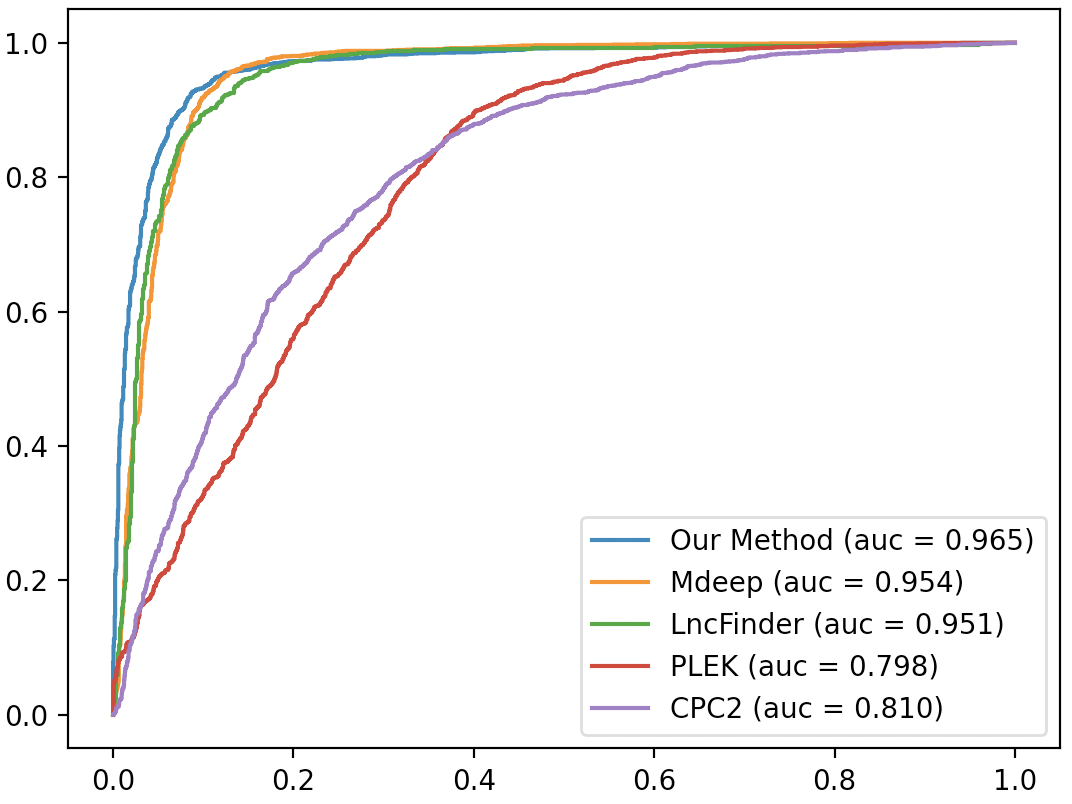}
        \label{fig:mo_80}
    \end{subfigure}
    \\
    \begin{subfigure}{0.4\textwidth}
        \centering
        \caption{} 
        \includegraphics[width=\textwidth]{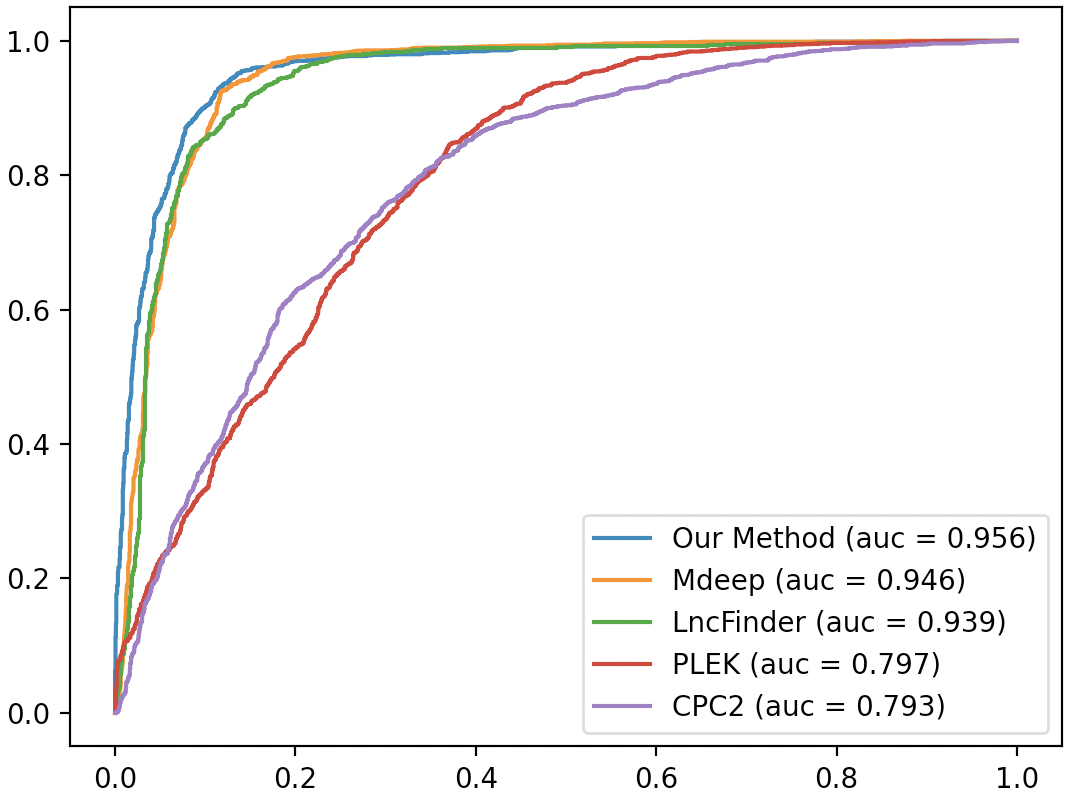}
        \label{fig:mo_70}
    \end{subfigure}
    \begin{subfigure}{0.4\textwidth}
        \centering
        \caption{} 
        \includegraphics[width=\textwidth]{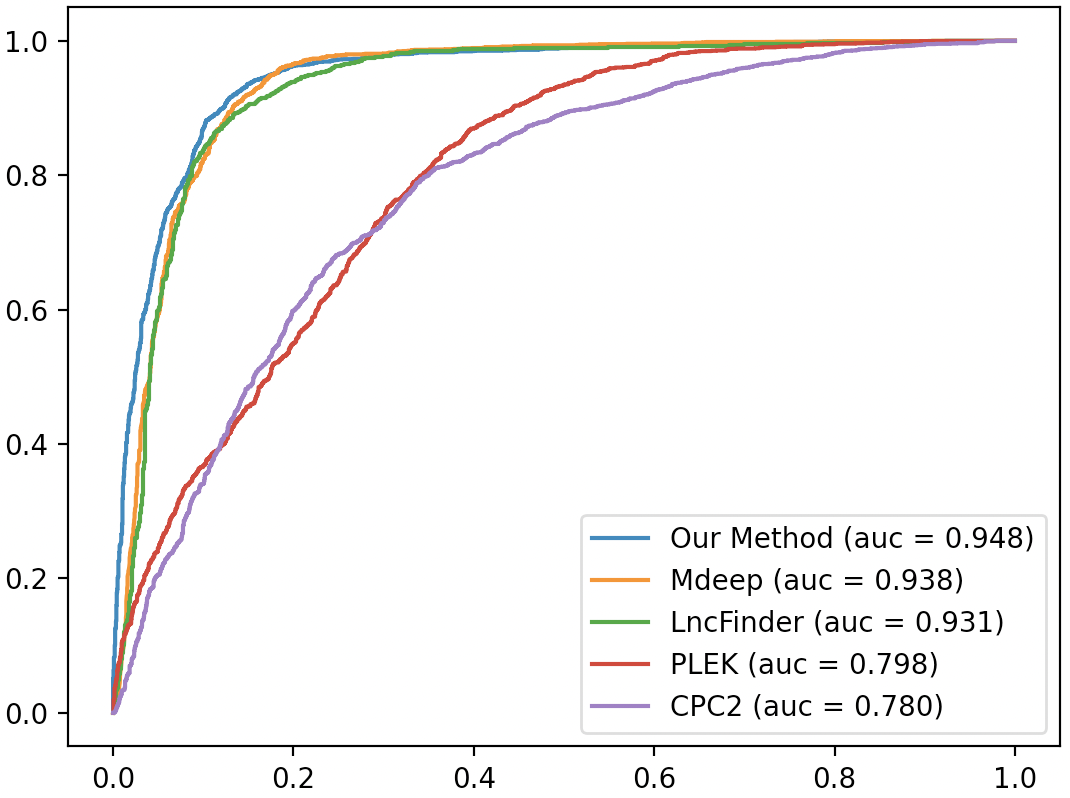}
        \label{fig:mo_60}
    \end{subfigure}
        \caption{The comparison with different models for ROC curve for different partial ratios for mouse species. (a) represent 90\% of partial sequence ROC curve, (b) represent 80\% of partial sequence ROC, (c) represent 70\% of partial sequence ROC curve, and (d) represent 60\% of partial sequence ROC curve.}
        \label{fig:mo_partial_roc}
\end{figure*}

We also compared the performance of our model with the same baseline methods. As results from the comparison in Table~\ref{table:hu_comaparison} and Figure~\ref{fig:hu_partial_roc} for a human transcript, our model performed better than other methods in precision and auROC values. For the accuracy and f1-score, mostly the Mdeep model had the highest performance over all different partial sequence length but our model also follows similar results as the Mdeep model, the difference was very low. Likewise, the results of the mouse transcript are shown in Table~\ref{table:mo_comaparison} and Figure~\ref{fig:mo_partial_roc}. In this case, our model performed the highest results for precision, accuracy, f1-score, and auROC. For the partial-sequence, our model successfully predicts the lncRNA and mRNA by implementing the ORF feature. Even though the sequence has been cut off with some portion, extracting the longest ORF from the remaining was successfully able to predict the sequence is lncRNA or mRNA.

\begin{table*}
    \centering
        \begin{threeparttable}
          \caption{Classification performance for human species for partial-sequence transcript.}
	\label{table:hu_comaparison}
	\scriptsize
	\begin{tabularx}{\linewidth}{XXXXXXX}
		\toprule
		\textbf{Partial} & \textbf{Methods} & \textbf{Precision} & \textbf{Recall} & \textbf{Accuracy} & \textbf{F1-score} & \textbf{auROC} \\
		\midrule
		& LncFinder & 0.771& 0.942& 0.830& 0.848& 0.854\\
            &Mdeep & 0.899& \textbf{0.948}& \textbf{0.920}& \textbf{0.922}& 0.960\\ 
90\% length &PLEK & 0.696& 0.939& 0.758& 0.780& 0.823\\ 
            &CPC2 & 0.627& 0.933& 0.687& 0.750& 0.787\\ 
            &\textbf{LoRA-BERT} & \textbf{0.926}& 0.902& 0.914& 0.914& \textbf{0.964}\\ 
		\midrule
		& LncFinder & 0.751& 0.946& 0.815& 0.837& 0.840\\
            &Mdeep & 0.885& 0.948& \textbf{0.912}& \textbf{0.915}& 0.955\\ 
80\% length &PLEK & 0.681& \textbf{0.953}& 0.745& 0.795& 0.820\\ 
            &CPC2 & 0.611& 0.940& 0.669& 0.741& 0.774\\ 
            &\textbf{LoRA-BERT} & \textbf{0.909}& 0.907& 0.908& 0.908& \textbf{0.958}\\ 
		\midrule
		& LncFinder & 0.729& 0.950& 0.797& 0.825& 0.830\\
            &Mdeep & 0.874& 0.946& \textbf{0.904}& \textbf{0.909}& 0.948\\ 
70\% length &PLEK & 0.662& \textbf{0.967}& 0.726& 0.786& 0.819\\ 
            &CPC2 & 0.599& 0.951& 0.653& 0.735& 0.766\\ 
            &\textbf{LoRA-BERT} & \textbf{0.897}& 0.912& 0.903& 0.904& \textbf{0.954}\\ 
		\midrule
		& LncFinder & 0.711& 0.957& 0.781& 0.816& 0.819\\
            &Mdeep & 0.858& 0.950& \textbf{0.895}& \textbf{0.902}& 0.940\\ 
60\% length &PLEK & 0.647& \textbf{0.977}& 0.706& 0.778& 0.822\\ 
            &CPC2 & 0.583& 0.956& 0.630& 0.724& 0.757\\ 
            &\textbf{LoRA-BERT} & \textbf{0.872}& 0.923& 0.893& 0.897& \textbf{0.941}\\ 
		\bottomrule
	\end{tabularx}
  \end{threeparttable}
\end{table*}

\begin{table*}
    \centering
        \begin{threeparttable}
          \caption{Classification performance for mouse species for partial-sequence transcript.}
	\label{table:mo_comaparison}
	\scriptsize
	\begin{tabularx}{\linewidth}{XXXXXXX}
		\toprule
		\textbf{Partial} & \textbf{Methods} & \textbf{Precision} & \textbf{Recall} & \textbf{Accuracy} & \textbf{F1-score} & \textbf{auROC} \\
		\midrule
		& LncFinder & 0.866& 0.952& 0.902& 0.907& 0.956\\
            &Mdeep & 0.914& 0.905& 0.910& 0.910& 0.959\\ 
90\% length &PLEK & 0.691& 0.918& 0.750& 0.789& 0.800\\ 
            &CPC2 & 0.637& \textbf{0.938}& 0.700& 0.759& 0.817\\ 
            &\textbf{LoRA-BERT} & \textbf{0.921}& 0.930& \textbf{0.925}& \textbf{0.926}& \textbf{0.967}\\ 
		\midrule
		& LncFinder & 0.851& 0.959& 0.895& 0.902& 0.951\\
            &Mdeep & \textbf{0.905}& 0.915& 0.908& 0.910& 0.954\\ 
80\% length &PLEK & 0.675& 0.939& 0.737& 0.785& 0.798\\ 
            &CPC2 & 0.619& \textbf{0.944}& 0.680& 0.748& 0.810\\ 
            &\textbf{LoRA-BERT} & 0.904& 0.933& \textbf{0.916}& \textbf{0.918}& \textbf{0.965}\\
		\midrule
		& LncFinder & 0.828& \textbf{0.955}& 0.877& 0.887& 
            0.939\\
            &Mdeep & \textbf{0.889}& 0.921& 0.902& 0.905& 0.946\\ 
70\% length &PLEK & 0.655& 0.954& 0.718& 0.777& 0.797\\ 
            &CPC2 & 0.603& 0.950& 0.659& 0.738& 0.793\\ 
            &\textbf{LoRA-BERT} & 0.883& 0.940& \textbf{0.907}& \textbf{0.911}& \textbf{0.956}\\
		\midrule
		& LncFinder & 0.808& 0.954& 0.862& 0.875& 0.931\\
            &Mdeep & \textbf{0.861}& 0.921& 0.885& 0.890& 0.938\\ 
60\% length &PLEK & 0.636& \textbf{0.970}& 0.696& 0.769& 0.798\\ 
            &CPC2 & 0.585& 0.958& 0.636& 0.727& 0.780\\ 
            &\textbf{LoRA-BERT} & 0.850& 0.945& \textbf{0.889}& \textbf{0.895}& \textbf{0.948}\\
		\bottomrule
	\end{tabularx}
  \end{threeparttable}
\end{table*}

\subsection{Sensitivity Analysis}

\begin{table*}
    \centering
        \begin{threeparttable}
          \caption{Classification performance for 90\% partial-sequence transcript without ORF feature and our original model.}
	\label{table:ORF}
	\scriptsize
	\begin{tabularx}{\linewidth}{XXXXXXX}
		\toprule
		\textbf{Dataset} & \textbf{Methods} & \textbf{Precision} & \textbf{Recall} & \textbf{Accuracy} & \textbf{F1-score} & \textbf{auROC} \\
		\midrule
		\multirow{2}{*}{Human}& Original & \textbf{0.926}&           0.902& \textbf{0.914}& \textbf{0.914}& \textbf{0.964}\\
            &No ORF & 0.604& \textbf{0.986}& 0.668& 0.749& 0.687\\ 
		\midrule
		\multirow{2}{*}{Mouse} & Original & \textbf{0.921}& 0.930& \textbf{0.925}& \textbf{0.926}& \textbf{0.967}\\
            &No ORF & 0.596& \textbf{0.991}& 0.658& 0.744& 0.717\\ 
		\bottomrule
	\end{tabularx}
  \end{threeparttable}
\end{table*}

The prediction performance of the partial-sequence transcript in our model is heavily reliant on the ORF feature. To validate the integrity of the ORF stage, we reconstructed the entire framework using solely the $k$-mer feature as the feature extraction and assessed the model's accuracy. We examined this modification using the human and mouse datasets and recorded the conclusive outcomes in Table~\ref{table:ORF}.

In Table~\ref{table:ORF}, you can find the performance metrics for both the original model and the reconstructed model that removed the ORF feature in partial-sequence identification. As anticipated, the reconstructed model was unable to make predictions for lncRNAs and mRNAs. Consequently, we can infer that the ORF feature is crucial in the context of LoRA-BERT because it has the capacity to capture the importance of a genomic segment in relation to amino acid composition or the signal for protein synthesis termination.

\section{Discussion}
State-of-the-art OMICs and bioinformatics analyses have revealed biological and physiological roles of lncRNAs. Despite these advances, there is lack of detailed characterization of the genetic regulation and functional features of lncRNAs, justifying the need for advanced approaches to identify and characterize lncRNA.

Transformers-based models have consistently delivered high performance across a range of NLP tasks, including biomedical and clinical entity extraction from extensive biomedical documents. In this context, we have introduced a BERT-based model with $k$-mer and ORF features for the specific identification of lncRNAs and mRNAs, relying solely on RNA sequences. Through a novel approach that involves creating global contextual embeddings for input sequences, we approach the task of predicting sequence specificity through a "BertForSequenceClassification" model. This approach commences by establishing a foundation of general RNA sequence supervised learning during the training process. We evaluated the model using two categories of datasets, one from humans and another from mouse and subsequently conducted a comparison with LncFinder, Mdeep, PLEK, and CPC2. The outcomes demonstrated that LoRA-BERT achieved an accuracy score of up to 98.5\% and the auROC is up to 99.8\% in the full-sequence human transcript, which indicates that our model is better than all the baselines above. Likewise in the partial-sequence transcript, our model holds stable results and state-of-the-art performance.

Through a sequence of discussions and experiments, our model, which utilizes feature extraction, has exhibited robust classification performance, enabling it to classify lncRNAs and mRNAs effectively. We believe this LoRA-BERT holds the potential to assist in the comprehension and detection of diseases associated with longer sequences, thereby enhancing our understanding of the broader biological processes. Furthermore, our research underscores the substantial promise of NLP models in the field of biosequence analysis.





\bibliographystyle{unsrt}
\bibliography{LoRABERT}  






\end{document}